\documentclass[conference]{IEEEtran}

\usepackage{times,amsmath,color,amssymb,graphicx,epsfig,cite,psfrag,enumerate,footmisc}

\long\def\symbolfootnote[#1]#2{\begingroup \def\thefootnote{\fnsymbol{footnote}}\footnote[#1]{#2}\endgroup}

\newtheorem{Remark}{\it Remark}[section]

\newtheorem{Proposition}{\it Proposition}[section]

\newtheorem{Corollary}{\it Corollary}[section]
\newtheorem{Definition}{\it Definition}[section]

\IEEEoverridecommandlockouts
\begin{document}

\title{On the Capacity of a Class of Cognitive Z-interference Channels}

\author{
\authorblockN{Jinhua Jiang$^{\dag}$, Ivana Mari\'{c}$^{\dag}$, Andrea Goldsmith$^{\dag}$, Shlomo Shamai (Shitz)$^{\ddag}$, and Shuguang Cui$^{\dag\dag}$\\}
\authorblockA{$^{\dag}$Dept. of EE, Stanford University, Stanford, CA 94305, USA\\
$^{\ddag}$Dept. of EE, Technion, Technion City, Haifa 32000, Israel\\
$^{\dag\dag}$Dept. of ECE, Texas A\&M University, College Station, TX 77843, USA\\
Email: jhjiang@stanford.edu, \{ivanam, andrea\}@wsl.stanford.edu, sshlomo@ee.technion.ac.il, and cui@ece.tamu.edu
}
%Jinhua Jiang$^{\dag}$, Ivana Mari\'{c}$^{\dag}$, Andrea Goldsmith$^{\dag}$, Shlomo Shamai (Shitz)$^{\ddag}$, and Shuguang Cui$^{\dag\dag}$
%\thanks{$^{\dag}$Dept. of EE, Stanford University, Stanford, CA 94305, USA}
%\thanks{$^{\ddag}$Dept. of EE, Technion, Technion City, Haifa 32000, Israel}
%\thanks{$^{\dag\dag}$Dept. of ECE, Texas A\&M University, College Station, TX 77843, USA}
%\thanks{Email: jhjiang@stanford.edu, \{ivanam, andrea\}@wsl.stanford.edu, sshlomo@ee.technion.ac.il, and cui@ece.tamu.edu}
\thanks{The work by J. Jiang, I. Mari\'{c}, and A. Goldsmith was supported by the DARPA ITMANET program under grant 1105741-1-TFIND. The work of S. Shamai was supported by the CORNET Consortium. The work by S. Cui was supported by the DTRA under grant HDTRA1-08-1-0010 and the AFOSR under grant FA9550-09-1-0107.}
}

%\author{
%\authorblockN{Jinhua Jiang, Ivana Mari\'{c}, Andrea Goldsmith\\}
%\authorblockA{Dept. of EE\\
%Stanford University \\
%Stanford, CA 94305, USA\\
%Email: jhjiang@stanford.edu, \{ivanam, andrea\}@wsl.stanford.edu}
%%\and\\
%\\
%\authorblockN{Shlomo Shamai (Shitz)\\}
%\authorblockA{Dept. of EE \\Technion  \\
%Technion City, Haifa 32000, Israel\\
%Email: sshlomo@ee.technion.ac.il}
%%\and\\
%\\
%\authorblockN{Shuguang Cui\\}
%\authorblockA{Dept. of ECE \\Texas A\&M University  \\
%College Station, TX 77843, USA\\
%Email: cui@ece.tamu.edu}}

% make the title area

\maketitle
%\linespread{1.0}

\begin{abstract}
%We study a special case of the cognitive radio channel, in which the receiver of the cognitive user does not suffer interference from the primary user. This corresponds to a class of cognitive `Z' channels that have not been investigated before. Conventionally, people would like to develop a coding scheme to cope with arbitrary channel conditions, but this kind of coding schemes usually result in rather cumbersome rate regions. However, in this paper, we focus on simple rate regions, which are easily computable. We first present a number of simple and explicit achievable rate region for the general discrete memoryless case. We then extend these rate regions to the Gaussian case. With a simple outer bound, we establish a new capacity result in the `high-interference' regime. Lastly, we also provide some numerical comparisons between the derived achievable rate regions and the outer bound.

%The numerical results indicate that the simplest coding scheme gives the largest achievable rate region, which is conjectured to the capacity region of the Gaussian cognitive `Z' channel in the high-interference regime.
We study a special class of the cognitive radio channel in which the receiver of the cognitive pair does not suffer interference from the  primary user. Previously developed general encoding schemes for this channel are complex as they attempt to cope with arbitrary channel conditions, which leads to rate regions that are difficult to evaluate. The focus of our work is to derive simple rate regions that are easily computable, thereby providing more insights into achievable rates and good coding strategies under different channel conditions. We first present several explicit achievable regions for the general discrete memoryless case. We also present an improved outer bound on the capacity region for the case of high interference. We then extend these regions to Gaussian channels. With a simple outer bound we establish a new capacity region in the high-interference regime. Lastly, we provide numerical comparisons between the derived achievable rate regions and the outer bounds.

%\footnotetext{The work by J. Jiang, I. Mari\'{c}, and A. Goldsmith was supported by the DARPA ITMANET program under grant 1105741-1-TFIND. The work of S. Shamai was supported by the CORNET Consortium. The work by S. Cui was supported by the DTRA under grant HDTRA1-08-1-0010 and the AFOSR under grant FA9550-09-1-0107.}

\end{abstract}

\section{Introduction}
Cognitive radio techniques bear the potential to significantly improve the efficiency of spectrum usage. As a result, the information-theoretic capacity gains associated with cognitive radios have been the subject of much investigation \cite{Tarokh06:ic_dms_cog,jovicic06:cog_ICDMS,wuwei06_icdms,Maric07:IC_partial_coop,Maric07:IC_cog_jnl,jiang07:ICDMS,Jiang09:BCCR_itw,Chenbiao09:IC_cog,Rini10:cog_ic}.
In these works, several achievable rate regions have been obtained by developing coding schemes based on rate splitting, Gel'fand-Pinsker coding (dirty paper coding), and superposition coding. The capacity regions for the general cases have not been determined except for the case of weak interference \cite{jovicic06:cog_ICDMS,wuwei06_icdms}, and the case of very strong interference \cite{Maric07:IC_partial_coop}. For a special case of the cognitive radio channel referred to as the cognitive Z-interference channel, the capacity region has been established under the assumption of a noiseless link from the primary transmitter to its receiver in \cite{liuNan08:cog_zic}. A more generalized model of the cognitive Z-interference channel has also been considered in \cite{caoyi09:cog_z}. Both of these works assume that the primary user's receiver does not suffer from interference generated by the cognitive user. Complementing the existing works, in this paper we consider another type of cognitive Z-interference channel, in which the receiver of the cognitive user does not suffer from interference generated by the primary user. This channel models the scenario where the link from the primary transmitter to the receiver of the cognitive user suffers from strong shadowing or other channel losses.

%, e.g., the cognitive pair and a TV receiver are in a room which is far from the primary user (the TV broadcasting station).

For this new type of cognitive Z-interference channel, achievable rate regions may be obtained by specializing some known achievable rate regions developed for the cognitive radio channel. As some of the coding schemes developed for the general case are intended to cope with arbitrary channel conditions, the associated achievable rate regions are very complex and hard to compute. As such, if we directly apply the general formulas to the considered Z-interference channel, the rate regions do not simplify despite the missing interference link. Instead, we will show in this paper that not all components of the general coding schemes are needed to achieve tight upper and lower bounds.

It has been shown that simple coding schemes are capacity achieving when the cognitive radio channel is in certain regimes. For example, the capacity region for the cognitive radio channel in the weak-interference regime is achievable with a very simple coding scheme \cite{jovicic06:cog_ICDMS,wuwei06_icdms}, and the capacity is defined by two inequalities. Similar phenomena can be found for the interference channel and the broadcast channel as well. In particular, the best achievable rate region for the general broadcast channel is Marton's region \cite{Marton79:BC_achievable}, while when the channel is Gaussian, the capacity region can be achievable by a simple superposition coding or a particular order of dirty paper coding. This is because an arbitrary Gaussian broadcast channel always falls into one of the two regimes, degraded or reversely degraded. For the interference channel, the best general coding scheme is the rate splitting scheme and the associated Han-Kobayashi rate region is quite complicated \cite{Han81:IFC}. However, when the interference is very weak, the capacity region is achievable by a naive coding scheme that simply ignores the interference; when the interference is strong, the capacity region is achievable by a simple scheme requiring the two receivers to decode both messages.

Therefore, in this paper, we look into simple coding schemes with simple achievable rate regions for the cognitive Z-interference channel. When the interference is weak, the capacity region can be obtained by directly extending the existing capacity result the cognitive radio channel in the weak-interference regime. Hence, we focus on the high-interference regime. We wish to examine the achievable rates associated with these simple coding schemes by comparing their achievable rate regions with the capacity region outer bound. Some of the simple coding schemes are in fact not new, but rather are extracted from the coding schemes previously proposed for various cognitive and interference channel models. Specifically, we first derive a few different simple achievable rate regions extended from the achievable rate region for the cognitive radio channel developed in \cite{Jiang09:BCCR_itw} and from a component code proposed in \cite{Tarokh06:ic_dms_cog,Maric07:IC_cog_jnl,jiang07:ICDMS}. We then extend these rate regions to the Gaussian case, and derive new outer bounds on the capacity region. We find that the latter simple achievable rate region is the capacity region, when the interference is in the lower range of the high-interference regime. Note that a similar capacity result and proof techniques were reported in an independent study on the Gaussian cognitive channel \cite{Rini10:cog_capacity}. In addition, our new outer bounds can be applied on the general cognitive channel in the high-interference regime.

%Through numerical comparisons, we also find that another simple achievable rate region that dominates all other simple achievable rate regions when the interference link is very strong.

The remainder of the paper is organized as follows. In Section II, we present the channel model and the related definitions. In Section III, we derive the aforementioned simple achievable rate regions and a new capacity region outer bound for the channel in the discrete memoryless case. In Section IV, we present the achievable rate regions and outer bounds for the Gaussian case, and a new capacity result in the high-interference regime, along with numerical comparisons between the achievable rate regions and the outer bounds. The paper is concluded in Section V.

\section{Channel Model}
\begin{figure}[h]
\centering
\psfrag{m1}{$m_1$}\psfrag{m2}{$m_2$}\psfrag{X1}{$X_1^{n}$}\psfrag{X2}{$X_2^n$}\psfrag{Y1}{$Y_1^n$}\psfrag{Y2}{$Y_2^n$}\psfrag{p}{$p(y_1|x_1)p(y_2|x_1,x_2)$}
\psfrag{m1e}{$\hat{m}_1$}\psfrag{m2e}{$\hat{m}_2$}
\includegraphics[width=0.6\linewidth]{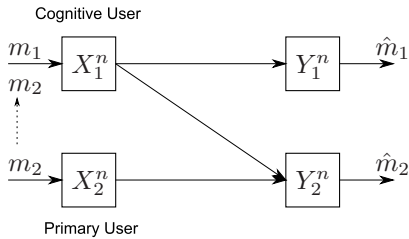}
\caption{Cognitive Z-interference channel.}\label{Channel_z_dmc}
\end{figure}

We consider the following channel model that involves one primary user and one cognitive user. As shown in Fig. \ref{Channel_z_dmc}, each user needs to send a message to its corresponding receiver, while the cognitive user is assumed to have non-causal knowledge of the primary user's message. The channel is defined by $(\mathcal{X}_1,\mathcal{X}_2,\mathcal{Y}_1,\mathcal{Y}_2,p(y_1,y_2|x_1,x_2)$, where $\mathcal{X}_t$ and $\mathcal{Y}_t$, $t=1,2$, denote the channel input and output alphabets, and $p(y_1,y_2|x_1,x_2)$ denotes the collection of channel transition probabilities. Furthermore, we assume $p(y_1,y_2|x_1,x_2)$ can be factored as $p(y_1|x_1)p(y_2|x_1,x_2)$, i.e., the cognitive user's receiver output is not affected by the primary user's channel input. The source messages $M_t$, $t=1,2$, are assumed to be uniformly generated over the respective ranges: $\mathcal{M}_t=\{1,2,...,\|\mathcal{M}_t\|\}$, $t=1,2$. We call this channel the cognitive Z-interference channel, and denote it as $\mathfrak{C}_{\mathrm{Z}}$.

%\begin{figure}[h]
%\centering
%\psfrag{m1}{$m_1$}\psfrag{m2}{$m_2$}\psfrag{X1}{$X_1^{n}$}\psfrag{X2}{$X_2^n$}\psfrag{Y1}{$Y_1^n$}\psfrag{Y2}{$Y_2^n$}
%\psfrag{m1e}{$\hat{m}_1$}\psfrag{m2e}{$\hat{m}_2$}
%\includegraphics[width=0.75\linewidth]{cog_ic_dmc.eps}
%\caption{Cognitive Z-interference channel.}\label{Channel_z_dmc}
%\end{figure}
%
%\begin{figure}[h]
%\centering
%\psfrag{m1}{$m_1$}\psfrag{m2}{$m_2$}\psfrag{X1}{$X_1^{n}$}\psfrag{X2}{$X_2^n$}\psfrag{Y1}{$Y_1^n$}\psfrag{Y2}{$Y_2^n$}
%\psfrag{b}{$b$}\psfrag{Z1}{$Z_1^n$}\psfrag{Z2}{$Z_2^n$}
%\includegraphics[width=0.75\linewidth]{cog_z_gaussian.eps}
%\caption{Cognitive Z-interference channel.}\label{Channel_z_dmc}
%\end{figure}

\begin{Definition}
An $(\|\mathcal{M}_1\|,\|\mathcal{M}_2\|,n,P_e^{(n)})$ code for $\mathfrak{C}_{\mathrm{Z}}$ consists of an encoding function at the cognitive user $ f_1:
\mathcal{M}_1 \times \mathcal{M}_2 \mapsto \mathcal{X}_1^n$, an encoding function at the primary user $f_2: \mathcal{M}_2 \mapsto \mathcal{X}_2^n$, and one decoding function at each receiver $g_t: \mathcal{Y}_t^n \mapsto \mathcal{M}_t$, $t=1,2$, with the probability of decoding errors defined as $P_e^{(n)} = \max\{P_{e,1}^{(n)},P_{e,2}^{(n)}\}$, where the individual error probability at each receiver is computed as $ P_{e,t}^{(n)} = \frac{1}{\|\mathcal{M}_1\|\|\mathcal{M}_2\|}\sum_{M_1\in \mathcal{M}_1,M_2\in \mathcal{M}_2}P(g_t(Y_t^n)\neq M_t|(M_1,M_2)~\mathrm{were~sent}), ~t=1,2. $ \label{Def_code}
\end{Definition}

\begin{Definition}
A non-negative rate pair $(R_1,R_2)$ is achievable for $\mathfrak{C}_{\mathrm{Z}}$ if there exists a sequence of codes $(2^{nR_1},2^{nR_2},n,P_e^{(n)})$
for the channel such that $P_e^{(n)}$ approaches 0 as $n\rightarrow \infty$. \label{Def_rate}
The capacity region of $\mathfrak{C}_{\mathrm{Z}}$ is the closure over the set of all achievable rate pairs. Any subset of the capacity region is an achievable rate region.
\end{Definition}

\section{Discrete Memoryless Channels}
In this section, we first derive a few simple achievable rate regions with explicit descriptions based on the rate region in \cite[Theorem 4.1]{Jiang09:BCCR_itw}, and with a component code applied in \cite{Tarokh06:ic_dms_cog,Maric07:IC_cog_jnl,jiang07:ICDMS}. We also derive a new outer bound on the capacity region of the channel in the high-interference regime.

\subsection{Achievable Rate Regions}
Let $U_1$, $W_1$, $V_2$, and $W_2$ be arbitrary auxiliary random variables defined over finite alphabets: $\mathcal{U}_1$, $\mathcal{W}_1$, $\mathcal{V}_2$, and $\mathcal{W}_2$, respectively. Let $\mathcal{P}$ denote the set of all joint distributions $p(\cdot)$ that factor in the following form
\begin{align*}
p(&u_1,w_1,v_2,w_2,x_1,x_2,y_1,y_2) \\
= &~ p(u_1)p(v_2)p(w_1,w_2|v_2,u_1) p(x_1|w_1,w_2,v_2,u_1)p(x_2|v_2)\\
&~ \cdot p(y_1|x_1)p(y_2|x_1,x_2)
\end{align*}
For any joint distribution $p(\cdot) \in \mathcal{P}$, define $\mathcal{R}(p(\cdot))$ as the set of non-negative rate pairs $(R_1,R_2)$ such that
\begin{align}
&R_1 \leq I(U_1,W_1;Y_1) - I(W_1;V_2|U_1),\\
&R_2 \leq I(V_2,W_2;Y_2|U_1),\\
&R_1 + R_2 \leq \min\{ I(V_2,W_2;Y_2|U_1) + I(U_1,W_1;Y_1)\notag\\
&~~~-I(W_1;W_2,V_2|U_1), I(V_2,W_2,U_1;Y_2) + I(W_1;Y_1|U_1) \notag\\
&~~~-I(W_1;W_2,V_2|U_1)\}.
%R_1 + 2R_2 \leq &~I(V_2,W_2,U_1;Y_2) + I(V_2,W_2;Y_2|U_1) + I(W_1;Y_1|U_1) - I(W_1;W_2,V_2|U_1),\\
%2R_1 + R_2 \leq &~I(W_1;Y_1|U_1) + I(V_2,W_2,U_1;Y_2) + I(U_1,W_1;Y_1) \notag \\&- I(W_1;V_2|U_1) - I(W_1;W_2,V_2|U_1),
\end{align}
and define $\mathcal{R} \triangleq \bigcup_{p(\cdot) \in \mathcal{P}}\mathcal{R}(p(\cdot))$.

\begin{Proposition}
The rate region $\mathcal{R}$ is an achievable rate region for the  cognitive Z-interference channel $\mathfrak{C}_{\mathrm{Z}}$.
\end{Proposition}

%\begin{proof}
{\it Proof}: The details of the proof is omitted here due to space limit. The proof is based on \cite[Theorem 4.1]{Jiang09:BCCR_itw}. First, note that the positions of the primary user and cognitive user are switched relative to the channel model in \cite{Jiang09:BCCR_itw}, and thus we switch the indices in the region description. Next, we drop the common information of primary user by setting the corresponding rate to be 0. Lastly, perform Fourier-Motzkin elimination on the implicit rate region, which obtains the set of inequalities defining $\mathcal{R}(p(\cdot))$. \hfill$\blacksquare$

%\end{proof}

\begin{Remark}
In this coding scheme, rate splitting is only applied on the cognitive user's message, i.e., $m_1 = (m_{12},m_{11})$. More specifically, the common message $m_{12}$ is encoded with codewords generated with $U_1$, and the private message $m_{11}$ is encoded with codewords generated with $W_1$. The primary user's message $m_2$ is encoded with codewords generated with both $V_2$ and $W_2$. The cognitive user's receiver decodes its own messages $m_1 = (m_{12},m_{11})$ only, while the primary user's receiver jointly decodes its own message $m_2$ and the common message from the cognitive user $m_{12}$. Next, we specialize the rate region $\mathcal{R}$ into two more simple achievable rate regions.
\end{Remark}

%\begin{Remark}
%Note that the rate region developed in \cite[Theorem 4.1]{Jiang09:BCCR_itw} is also an achievable rate region for the cognitive `Z' channel under investigation, which appears to be more general. Nevertheless, we choose to focus on a sub region of this rate region for two main reasons: 1) ; 2).
%\end{Remark}

First, we completely remove the `artificially' created common information, i.e., no rate splitting is applied to the cognitive user's message. The following rate region can be obtained by setting $U_1$ as a constant.

\begin{Corollary}
Any rate pair $(R_1,R_2)$ satisfying
\begin{align}
R_1 \leq &~I(W_1;Y_1) - I(W_1;V_2),\label{region_R1_1}\\
R_2 \leq &~I(V_2,W_2;Y_2),\\
R_1 + R_2 \leq &~ I(V_2,W_2;Y_2) + I(W_1;Y_1)-I(W_1;W_2,V_2),\label{region_R1_3}
\end{align}
for any joint distribution
\begin{align}
p(&w_1,v_2,w_2,x_1,x_2,y_1,y_2) =  p(v_2)p(w_1,w_2|v_2) \notag\\
&~  \cdot p(x_1|w_1,w_2,v_2)p(x_2|v_2) p(y_1|x_1)p(y_2|x_1,x_2),
\end{align}
is achievable for the channel $\mathfrak{C}_{\mathrm{Z}}$.
\end{Corollary}
\begin{Remark}
Denote the rate region \eqref{region_R1_1}--\eqref{region_R1_3} as $\mathcal{R}_1$. Note that $W_1$ and $W_2$ are used to perform Marton's binning. Therefore, when it is extended to the Gaussian case, it becomes two different rate regions as a result of dirty paper coding in different orders.
\end{Remark}

For the second one, instead of removing the common information of the cognitive user, we remove the private information. Specifically, set $W_1$ as a constant, and merge $V_2$ and $W_2$ into one random variable $V_2$, i.e., set $W_2=V_2$. We have the following simple achievable rate region.

\begin{Corollary}
Any rate pair $(R_1,R_2)$ satisfying
\begin{align}
R_1 \leq &~I(U_1;Y_1),\label{Region_R2_1}\\
R_2 \leq &~I(V_2;Y_2|U_1),\\
R_1 + R_2 \leq &~I(V_2,U_1;Y_2),\label{Region_R2_3}
\end{align}
for any joint distribution
\begin{align}
p(u_1,v_2,x_1,x_2,y_1,y_2) = &~ p(u_1)p(v_2) p(x_1|v_2,u_1)p(x_2|v_2) \notag\\&~ \cdot p(y_1|x_1)p(y_2|x_1,x_2),
\end{align}
is achievable for the channel $\mathfrak{C}_{\mathrm{Z}}$.
\end{Corollary}

\begin{Remark}
Denote the rate region \eqref{Region_R2_1}--\eqref{Region_R2_3} as $\mathcal{R}_2$. This rate region is achieved by using only superposition coding. Nevertheless, it turns out to be the largest amongst all these simple rate regions when the interference link is `substantially' strong, which is numerically shown in the next section.
\end{Remark}

We now present another simple achievable rate region, based on the idea that the artificial common message of the cognitive user can be dirty-paper coded against the primary user's private message. The purpose of dirty paper coding in this coding scheme is to protect the message of the cognitive user from the interference created by the primary user's message during the cognitive user's encoding process.

\begin{Proposition}
Any rate pair $(R_1,R_2)$ satisfying
\begin{align}
R_1 \leq &~ I(U_1;Y_1) - I(U_1;V_2),
\\
R_2 \leq&~ I(V_2;U_1,Y_2),
\\
R_1 + R_2 \leq& ~ I(U_1,V_2;Y_2),
\end{align}
for any joint distribution
\begin{align}
p(u_1,v_2,x_1,x_2,y_1,y_2) = &~ p(u_1,v_2) p(x_1|v_2,u_1)p(x_2|v_2) \notag\\&~ \cdot p(y_1|x_1)p(y_2|x_1,x_2),
\end{align}
is achievable for the channel $\mathfrak{C}_{\mathrm{Z}}$.
\end{Proposition}
\begin{Remark}
Let us denote this rate region by $\mathcal{R}_3$. The proof can be easily obtained by following the argument in \cite{jiang07:ICDMS}. The intention of the dirty paper coding applied in this coding scheme is to let the encoding of the cognitive user's common message take advantage of its non-causal knowledge about the primary user's private message. As a component of the respective coding schemes, this technique has been employed in several existing works \cite{Tarokh06:ic_dms_cog,Maric07:IC_cog_jnl,jiang07:ICDMS}. Nevertheless, there has been no evidence showing that this feature is indeed helpful in terms of increasing achievable rates. In the next section, we will show that this coding scheme is in fact capacity-achieving, when the channel is in the high-interference regime, but the interference belongs to the lower range.

%In two recent works \cite{Chenbiao09:IC_cog} and \cite{Jiang09:BCCR_itw}, such a coding feature is dropped. This is because, when two messages are required to be decoded at a common receiver, and one message is dirty paper coded against the other, the rate increment of one message is at the cost of a rate reduction of the other. This phenomenon is in fact implied by the findings in \cite{Kim08:state_amplification}. Nevertheless,
\end{Remark}

\subsection{A New Outer Bound}
In \cite[Theorem 5]{Maric07:IC_cog_jnl}, an outer bound on the capacity region of the strong cognitive interference channel was derived. This outer bound also directly applies to the cognitive Z-interference channel in the high-interference regime. It was also observed in \cite{Maric07:IC_cog_jnl} that, under the strong/high interference assumption, the primary user can decode the cognitive user's message without any rate penalty. In fact, an equivalence between the capacity regions of the strong cognitive interference channel and its variant with degraded message sets can be established. The capacity region of strong cognitive interference channel can now be outer-bounded by the capacity region of the broadcast channel with degraded message sets \cite{korner77:bc_degraded_message_set}, where we let the primary user and the cognitive user fully cooperate. By combining this with the outer bound derived in \cite[Theorem 5]{Maric07:IC_cog_jnl}, we obtain a new outer bound for the cognitive interference channel in the high-interference regime. Denote the outer bound given in \cite[Theorem 5]{Maric07:IC_cog_jnl} as $\mathcal{C}^o_1$.

%the set of all rate pairs satisfying
%\begin{align}
%R_1 &\leq I(X_1;Y_1|X_2),\\
%R_1+ R_2 & \leq I(X_1,X_2;Y_2),
%\end{align}
%for all joint input distributions $p(x_1,x_2)$.
Define $\mathcal{C}^o_2$ as the set of all rate pairs satisfying
\begin{align}
R_1 &\leq \min \{I(X_1;Y_1|X_2),I(U; Y_1)\},\\
%R_1 &\leq  I(U; Y_1),\\
R_2 &\leq I(X_1,X_2; Y_2|U),\\
R_1+R_2 &\leq I(X_1,X_2;Y_2),
\end{align}
for all joint input distributions $p(u,x_1,x_2)$.

\begin{Proposition}
For a cognitive Z-interference channel satisfying $I(X_1;Y_1|X_2) \leq I(X_1;Y_2|X_2)$
for all input distributions $p(x_1,x_2)$, $\mathcal{C}^o_2$ is an outer bound on the capacity region.
\end{Proposition}

\begin{Remark}
In general, the new outer bound $\mathcal{C}^{o}_2$ can be interpreted as an intersection between $\mathcal{C}^{o}_1$ and capacity region of the associated broadcast channel with degraded message sets, and thus it is a subset of the outer bound $\mathcal{C}^{o}_1$. This outer bound is strictly smaller in some cases, which is demonstrated using a numerical example in the next section.
\end{Remark}

\section{Gaussian Channels}
In this section, we first define the Gaussian channel model. We then extend the achievable rate regions derived in the previous section to the Gaussian case, and present a new capacity result in the high-interference regime.

%\subsection{Channel model}
%The cognitive user and primary user are subject to transmit power constraint, $P_1$ and $P_2$, respectively. Channel gains of the links from the users to the respective intended receiver are assumed to be 1, and the only cross link gain (from the cognitive user to the receiver of primary user) is assumed to be of non-negative real value $b$. In addition, additive white Gaussian noises of zero mean and unit variance are assumed at both receivers.
The channel input-output relationship is given by
\begin{align}
Y_1 &= X_1 + Z_1, \label{channel_gaussian_1}\\
Y_2 &= X_2 + b X_1 + Z_2, \label{channel_gaussian_2}
\end{align}
where $b$ is the interference link gain and is assumed to be greater than $1$; $Z_1$ and $Z_2$ are additive white Gaussian noises of zero mean and unit variance. The power constraints are given by
$
\frac{1}{n}\sum_{i=1}^{n}\|X_t\|^2 \leq P_t, ~ t = 1,2
$.

Note that when $b \leq 1$, the channel belongs to the weak-interference regime, for which the capacity region has been established for the general cognitive radio channel \cite{jovicic06:cog_ICDMS,wuwei06_icdms}. The capacity achieving scheme is rather simple: the cognitive user spares a portion of its own power to cooperate with the primary user for the transmission of message $m_2$, while it performs dirty paper coding on its own message $m_1$ by treating the signals carrying $m_1$ as the non-causally known interference. No rate splitting is applied, and each receiver only decodes its own intended message. Here, we focus on the high-interference regime, which is in general open. In what follows, we present the Gaussian counterparts of the achievable regions presented in the previous section.

\subsection{Achievable Rate Regions}
We first extend the rate region $\mathcal{R}$ to the Gaussian case. When Marton's coding scheme is extended to the Gaussian case, the double binning first extends
to Gel'fand-Pinsker coding in two different orders, and subsequently dirty-paper coding in two different orders. The rate region $\mathcal{R}$ can thus be represented as a union of two rate regions corresponding to the respective dirty-paper coding orders. Define
$\mathcal{G}_a(\alpha,\beta,\theta)$ as the set of rate pairs satisfying:
\begin{align}
R_1 &\leq \frac{1}{2}\log_2\left(1+ \frac{\alpha\beta P_1}{\alpha\bar{\beta}P_1+\bar{\alpha}P_1+1}\right) \notag\\&~~~~+ \frac{1}{2}\log_2\left(1+\frac{\alpha\bar{\beta}P_1}{\bar{\alpha}\bar{\theta}P_1+1}\right),\\
R_2 &\leq \frac{1}{2}\log_2\left(1+\frac{(\sqrt{P_2}+b\sqrt{\bar{\alpha}\theta P_1})^2}{b^2\alpha\bar{\beta}P_1+b^2\bar{\alpha}\bar{\theta}P_1+1}\right)  \notag\\&~~~~+ \frac{1}{2}\log_2\left(1+b^2\bar{\alpha}\bar{\theta}P_1\right),
\\
R_1 + R_2 &\leq  \frac{1}{2}\log_2\left(1+\frac{(\sqrt{P_2}+b\sqrt{\bar{\alpha}\theta P_1})^2+b^2\alpha\beta P_1}{b^2\alpha\bar{\beta}P_1+b^2\bar{\alpha}\bar{\theta}P_1+1}\right)\notag\\&~~~~+\frac{1}{2}\log_2\left(1+\frac{\alpha\bar{\beta}P_1}{\bar{\alpha}\bar{\theta}P_1+1}\right) + \frac{1}{2}\log_2\left(1+b^2\bar{\alpha}\bar{\theta}P_1\right),
\end{align}
and  $\mathcal{G}_b(\alpha,\beta)$ as the set of rate pairs satisfying:
\begin{align}
% R_1 & \leq \frac{1}{2}\log_2\left(\right),\\
R_1 & \leq \frac{1}{2}\log_2\left(1+ \frac{\alpha \beta P_1}{1+ \alpha \bar{\beta}P_1 + \bar{\alpha}P_1}\right) \notag\\&~~~~+  \frac{1}{2}\log_2\left(1+ \alpha \bar{\beta} P_1\right),\\
R_2 & \leq \frac{1}{2}\log_2\left(1+ \frac{(b\sqrt{\bar{\alpha}P_1}+\sqrt{P_2})^2}{1+b^2\alpha\bar{\beta}P_1}\right),\\
R_1 + R_2 & \leq \frac{1}{2}\log_2\left(1+\frac{(b\sqrt{\bar{\alpha}P_1}+\sqrt{P_2})^2+b^2\alpha\beta P_1}{1+b^2\alpha\bar{\beta}P_1}\right) \notag\\&~~~~+  \frac{1}{2}\log_2\left(1+\alpha\bar{\beta}P_1\right),
\end{align}
where $0 \leq \alpha \leq 1$, $0 \leq \beta \leq 1$, and $0 \leq \theta \leq 1$; and $\bar{x} \triangleq 1-x$. Next, define $\mathcal{G} \triangleq \mathrm{conv} \{ \bigcup_{\alpha,\beta,\theta}\mathcal{G}_b(\alpha,\beta,\theta)\cup\mathcal{G}_b(\alpha,\beta)\}$, where $\mathrm{conv}$ stands for the convex hull operation.

\begin{Corollary}
Any rate pair $(R_1,R_2)\in \mathcal{G}$ is achievable for the Gaussian Z-interference channel.
\end{Corollary}

Next, we extend the rate region $\mathcal{R}_1$ to the Gaussian case. Note that $\mathcal{R}_1$ is extended from $\mathcal{R}$ by setting $U_1$ as a constant. Hence, it is straightforward to obtain the Gaussian counterpart of $\mathcal{R}_1$ by setting $\beta=0$ which corresponds to a constant $U_1$. This Gaussian rate region can be expressed as $\mathcal{G}_1 \triangleq \mathrm{conv} \{ \bigcup_{\alpha,\beta=0,\theta}\mathcal{G}_b(\alpha,\beta,\theta)\cup\mathcal{G}_b(\alpha,\beta)\}$.

We also extend $\mathcal{R}_2$ to the its Gaussian counterpart $\mathcal{G}_2$, which is defined as the set of rate pairs satisfying:
%as follows. Let
%\begin{align}
%X_1 &= \sqrt{\alpha P_1} U_1 + \sqrt{\bar{\alpha} P_1} V_2,\\
%X_2 &= \sqrt{P_2} V_2,
%\end{align}
%and we easily find $\mathcal{G}_2$ defined by the following rate constraints
\begin{align}
R_1 &\leq \frac{1}{2}\log_2\left(1+ \frac{\alpha P_1}{1+\bar{\alpha} P_1}\right),\\
R_2 & \leq \frac{1}{2}\log_2\left(1+ (b\sqrt{\bar{\alpha}P_1}+\sqrt{P_2})^2\right),\\
R_1 + R_2 &\leq \frac{1}{2}\log_2\left(1+(b\sqrt{\bar{\alpha}P_1}+\sqrt{P_2})^2 + b^2 \alpha P_1\right).
\end{align}

Lastly, we extend $\mathcal{R}_3$ to obtain the Gaussian rate region $\mathcal{G}_3$, and we have
%The following mapping of random variables is applied
%\begin{align}
%U_1 & = \tilde{U}_1 + \lambda V_2,\\
%X_1 & = \sqrt{\alpha P_1} \tilde{U}_1 + \sqrt{\bar{\alpha}P_1}V_2,\\
%X_2 & = \sqrt{P_2}V_2.
%\end{align}
%Evaluating the rate constraints defining $\mathcal{R}_3$ with respect to the above mapping, we obtain the rate region $\mathcal{G}_3$ defined by the following rate constraints
\begin{align}
&R_1  \leq \frac{1}{2}\log_2\left(\frac{P_1 + 1}{(1+\lambda^2)(P_1 + 1)-(\sqrt{\alpha P_1}+\lambda\sqrt{\bar{\alpha}P_1})^2}\right), \label{Region_G3_1}\\
&R_2  \leq \frac{1}{2}\log_2\bigg((1+\lambda^2)((b\sqrt{\bar{\alpha} P_1}+\sqrt{P_2})^2+b^2\alpha P_1+1) \notag\\&~~~~~~~-(b\sqrt{\alpha P_1}+\lambda(\sqrt{P_2}+b\sqrt{\bar{\alpha} P_1}))^2\bigg),\\
&R_1+R_2  \leq \frac{1}{2}\log_2\left((b\sqrt{\bar{\alpha} P_1}+\sqrt{P_2})^2+b^2\alpha P_1+1\right).
\end{align}

By choosing $\lambda = \lambda_o = \sqrt{\alpha\bar{\alpha}}P_1/(\alpha P_1 +1)$, which is the dirty-paper coding coefficient maximizing the righthand side of \eqref{Region_G3_1}, we have another achievable rate region  $\mathcal{G}_3^{'}$ defined by the following rate constraints
\begin{align}
&R_1  \leq \frac{1}{2}\log_2\left(1+\alpha P_1\right), \label{Region_G3_11}
\end{align}
\begin{align}
&R_2  \leq \frac{1}{2}\log_2\bigg((1+\lambda_o^2)((b\sqrt{\bar{\alpha} P_1}+\sqrt{P_2})^2+b^2\alpha P_1+1)\notag\\&~~~~~~~-(b\sqrt{\alpha P_1}+\lambda_o(\sqrt{P_2}+b\sqrt{\bar{\alpha} P_1}))^2\bigg),\label{Region_G3_22}\\
& R_1+R_2  \leq \frac{1}{2}\log_2\left((b\sqrt{\bar{\alpha} P_1}+\sqrt{P_2})^2+b^2\alpha P_1+1\right). \label{Region_G3_33}
\end{align}

\begin{Remark}
It is easy to see that $\mathcal{G}_3^{'} \subseteq \mathcal{G}_3$. However, the rate region $\mathcal{G}_3$ cannot be easily computed, due to the fact that the choice of $\lambda$ is unbounded, i.e., $\lambda \in [0, +\infty)$, while $\mathcal{G}_3^{'}$ can be easily computed. Furthermore, the rate region $\mathcal{G}_3^{'}$ even turns out to be the capacity region when interference link gain falls into a certain range in the high-interference regime, as shown in Section IV.C.
\end{Remark}

\subsection{Outer Bounds}
As we assume that the cross link gain $b$ is greater than 1, the channel satisfies the conditions for the capacity outer bound derived in \cite[Corollary 1]{Maric07:IC_cog_jnl}. Hence, this outer bound directly applies to the cognitive Z-interference channel in the high-interference regime.

\begin{Proposition}
Any achievable rate pair $(R_1,R_2)$ for the Gaussian cognitive Z-interference channel defined by \eqref{channel_gaussian_1}--\eqref{channel_gaussian_2} with $b\geq1$, satisfies the following constraints
\begin{align}
R_1 & \leq \frac{1}{2}\log_2\left(1+ (1-\rho^2)P_1\right),\label{outer_1}\\
R_1 + R_2 & \leq \frac{1}{2}\log_2\left(1+ b^2P_1+P_2+2\rho b\sqrt{P_1P_2}\right), \label{outer_2}
\end{align}
for any $0\leq \rho \leq 1$. \label{outer_bound_1}
\end{Proposition}

Our improved outer bound $\mathcal{C}^o_2$ for the Gaussian case can be obtained by intersecting the outer bound in Proposition \ref{outer_bound_1} with the capacity region of the associated Gaussian MISO broadcast channel with degraded message sets \cite{shamai06:bc_dms}. Note that the computation of the region  $\mathcal{C}^{\mathrm{BCDMS}}$ requires optimization with respect to the individual antenna power constraints.

\subsection{Capacity in the High-interference Regime}
We observe that the rate constraints \eqref{Region_G3_11} and \eqref{Region_G3_33} are equivalent to \eqref{outer_1} and \eqref{outer_2}, when we let $\rho = \sqrt{1-\alpha}$. Hence, the rate region $\mathcal{G}_3^{'}$ meets the outer bound if the rate constraint \eqref{Region_G3_22} is loose.

\begin{Proposition}
For the Gaussian cognitive Z-interference channel defined by \eqref{channel_gaussian_1}--\eqref{channel_gaussian_2}, the capacity region is defined by
\begin{align}
R_1  &\leq \frac{1}{2}\log_2\left(1+\alpha P_1\right),\\
R_1+R_2  &\leq \frac{1}{2}\log_2\left((b\sqrt{\bar{\alpha} P_1}+\sqrt{P_2})^2+b^2\alpha P_1+1\right)
\end{align}
with $0\leq\alpha\leq 1$, when the interference link gain satisfies \[1\leq b \leq \sqrt{\frac{P_1+P_2+1}{P_1+1}} .\]
\end{Proposition}
\begin{Remark}
The above proposition is proved by showing every rate pair on the boundary of the outer bound is achievable. For the rate constraint \eqref{Region_G3_22} to be loose, the righthand side of \eqref{Region_G3_22} has to be greater than or equal to the difference between the righthand sides of \eqref{Region_G3_33} and \eqref{Region_G3_11}. Through cumbersome but simple algebra, the condition can be found as \[b^2 \leq \frac{P_1 + P_2 + \alpha P_1 P_2 +1}{P_1+1}. \] For this condition to be satisfied for all the choices of $\alpha \in [0,1]$, we need  \[b^2 \leq \frac{P_1 + P_2 +1}{P_1+1}. \]
\end{Remark}

\subsection{Numerical Results}
In this subsection, we numerically compare the rate regions and the outer bound. For all the comparisons, we let $P_1 = P_2 = 6$, leaving the interference link gain $b$ to be variable.

\begin{figure}[t]
\centering
\includegraphics[width=0.78\linewidth]{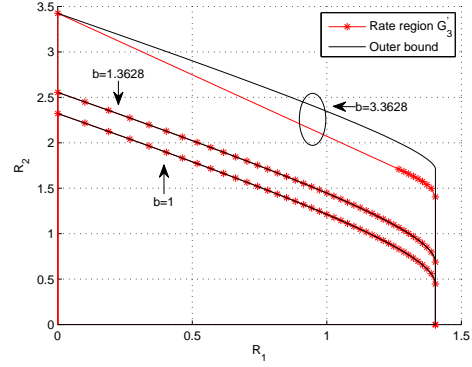}
\caption{Comparison between the rate region $\mathcal{G}_3^{'}$ and the outer bound $\mathcal{C}^{o}_1$.}\label{Fig_G3}
\end{figure}

First, in Fig. \ref{Fig_G3}, the rate region $\mathcal{G}_3^{'}$ is compared against the outer bound $\mathcal{C}^{o}_1$.  We observe from the figure that the achievable rate region meets the outer bound for both $b=1$ and $b = \sqrt{\frac{P_1+P_2+1}{P_1+1}}=1.3628$, while the rate region becomes strictly smaller than the outer bound for $b=3.3628>\sqrt{\frac{P_1+P_2+1}{P_1+1}}$. When the interference link becomes very strong, the coding scheme to achieve $\mathcal{G}_3^{'}$ turns out to be suboptimal, which is shown in the later comparisons.

\begin{figure}[t]
\centering
\includegraphics[width=0.78\linewidth]{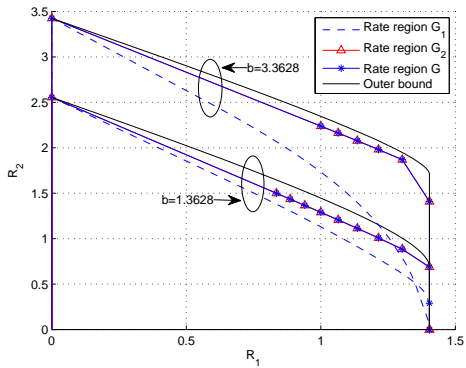}
\caption{Comparison of the rate regions $\mathcal{G}$, $\mathcal{G}_1$, $\mathcal{G}_2$, and the outer bound $\mathcal{C}^{o}_1$.}\label{Fig_G012}
\end{figure}

Next, in Fig. \ref{Fig_G012}, we compare the rate regions $\mathcal{G}$, $\mathcal{G}_1$, $\mathcal{G}_2$, and the outer bound $\mathcal{C}^{o}_1$ in the very-high interference regime, i.e., $b\geq \sqrt{\frac{P_1+P_2+1}{P_1+1}}=1.3628$. It can be observed that $\mathcal{G}_1$ is strictly smaller than $\mathcal{G}$ and $\mathcal{G}_2$, while $\mathcal{G}$ and $\mathcal{G}_2$ always coincide with each other. Note that $\mathcal{G}_2$ is in fact a subset of $\mathcal{G}$, as the coding scheme to achieve $\mathcal{G}$ generalizes the one for $\mathcal{G}_2$. However, the numerical results indicate that the coding scheme for $\mathcal{G}_2$ dominates other coding components in the coding scheme to achieve $\mathcal{G}$. This may imply that other coding features, especially the one to achieve $\mathcal{G}_1$, are redundant; this conjecture is the subject of further investigation.

\begin{figure}[t]
\centering
\includegraphics[width=0.78\linewidth]{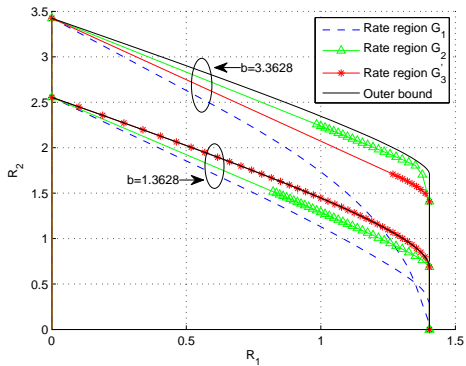}
\caption{Comparison of the rate regions $\mathcal{G}_1$, $\mathcal{G}_2$, $\mathcal{G}_3^{'}$, and the outer bound $\mathcal{C}^{o}_1$.}\label{Fig_G123}
\end{figure}

We also compare the rate regions $\mathcal{G}_1$, $\mathcal{G}_2$, $\mathcal{G}_3^{'}$, and the outer bound $\mathcal{C}^{o}_1$ in the very high-interference regime in Fig. \ref{Fig_G123}. At the boundary, when $b=1.3628$, the rate region $\mathcal{G}_3^{'}$ meets the outer bound, and is strictly larger than both $\mathcal{G}_1$ and $\mathcal{G}_2$. However, when $b$ is large enough, i.e., when $b=3.3628$, the rate region $\mathcal{G}_2$ becomes strictly larger than $\mathcal{G}_3^{'}$.  It is a current topic of investigation to find the critical value of $b$, at which $\mathcal{G}_2$ becomes larger than $\mathcal{G}_3^{'}$.

\begin{figure}[t]
\centering
\includegraphics[width=0.78\linewidth]{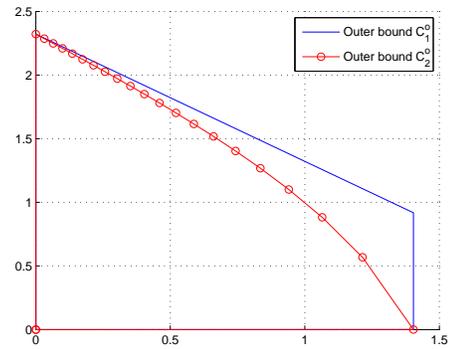}
\caption{Comparison of the outer bounds $\mathcal{C}^{o}_2$ and $\mathcal{C}^{o}_1$.}\label{Fig_outerbounds}
\end{figure}

Lastly, we compare the new outer bound $\mathcal{C}^{o}_2$ with $\mathcal{C}^{o}_1$ under an extreme case setting: $P_1=6$, $P_2=0$, and $b=2$. In Fig. \ref{Fig_outerbounds}, it can be easily observed that $\mathcal{C}^{o}_2$ is strictly smaller than $\mathcal{C}^{o}_1$.

% i.e., $\mathcal{C}^{o}_2$ provides a tighter outer bound on the capacity region of the cognitive Z-interference channel in the high-interference regime.

\section{Conclusions}
We have investigated the cognitive Z-interference channel where the cognitive user's receiver suffers no interference from the primary user. Our results demonstrate that simple coding schemes perform well in the strong-interference regime. In particular, we have shown that the capacity region is achieved with an encoding scheme that combines superposition coding and dirty paper coding (corresponding to $\mathcal{R}_3$)) over certain regime of interference, i.e.  when $1\leq b \leq \sqrt{\frac{P_1+P_2+1}{P_1+1}}$, Furthermore, our numerical results suggest that the superposition coding scheme (corresponding to $\mathcal{R}_2$) strictly dominates other schemes when interference is very strong. However, there is a gap between the achievable rate region and the considered outer bound. We also cannot show that the improved outer bound is tight. Therefore, further tightening of the outer bounds or deriving a new one is a
possible direction of the future work.

%In this paper, we have investigated a new class of cognitive `Z' channel. We did not try to develop a coding scheme to cope with arbitrary channel conditions. Instead, we focused on simple coding schemes and achievable rate regions for the channel. Our findings indicate that simple coding schemes in fact perform very well for this cognitive `Z' channel, especially in the high-interference regime. On one hand, we have shown that the capacity region can be achieve with one simple coding scheme (corresponding to $\mathcal{R}_3$) when the interference is not too strong, i.e., $1\leq b \leq \sqrt{\frac{P_1+P_2+1}{P_1+1}}$. On the other hand, our numerical results suggest that another simple coding scheme (corresponding to $\mathcal{R}_2$) strictly dominates others when the interference is very strong. However, there is a gap between this rate region and the outer bound in this unidentified regime. We suspect that this gap is due to the looseness of the simple outer bound. It is an interesting future work to further investigate on this issue.

\bibliographystyle{IEEETran}
\bibliography{IEEEAbrv,D:/MyPapers/mybib/mybib}
\end{document}